\documentclass[twocolumn,showpacs,superscriptaddress,prb,aps,floatfix]{revtex4-2}
\usepackage{graphicx}
\usepackage{rotating}
\usepackage{amsmath}
\usepackage[usenames,dvipsnames]{color}
\usepackage{float}
\usepackage{rotating}
\usepackage{booktabs}
\usepackage{multirow}
\usepackage{longtable}
\usepackage[colorlinks,linkcolor=blue,citecolor=magenta]{hyperref}
\usepackage[T1]{fontenc}
\usepackage{bm}
\usepackage{upgreek}
\usepackage{comment}
\usepackage{amssymb}

\hfuzz=\maxdimen
\begin{document}
	
\title{Strong Intra- and Interchain Orbital Coupling Leads to Multiband and High Thermoelectric Performance in Na$_2$Au$X$ ($X$ = P, As, Sb, and Bi)}

\author{Zhonghao Xia}
\affiliation{Key Laboratory of Advanced Materials and Devices for Post-Moore Chips, Ministry of Education, University of Science and Technology Beijing, Beijing 100083, China}
\affiliation{Beijing Key Laboratory for Magneto-Photoelectrical Composite and Interface Science, School of Mathematics and Physics, University of Science and Technology Beijing, Beijing 100083, China}

\author{Zhilong Yang}
\affiliation{Key Laboratory of Advanced Materials and Devices for Post-Moore Chips, Ministry of Education, University of Science and Technology Beijing, Beijing 100083, China}
\affiliation{Beijing Key Laboratory for Magneto-Photoelectrical Composite and Interface Science, School of Mathematics and Physics, University of Science and Technology Beijing, Beijing 100083, China}

\author{Yali Yang}
\affiliation{Key Laboratory of Advanced Materials and Devices for Post-Moore Chips, Ministry of Education, University of Science and Technology Beijing, Beijing 100083, China}
\affiliation{Beijing Key Laboratory for Magneto-Photoelectrical Composite and Interface Science, School of Mathematics and Physics, University of Science and Technology Beijing, Beijing 100083, China}

\author{Kaile Ren}
\affiliation{Beijing Key Laboratory for Magneto-Photoelectrical Composite and Interface Science, School of Mathematics and Physics, University of Science and Technology Beijing, Beijing 100083, China}

\author{Jiangang He}
\email{jghe2021@ustb.edu.cn}
\affiliation{Key Laboratory of Advanced Materials and Devices for Post-Moore Chips, Ministry of Education, University of Science and Technology Beijing, Beijing 100083, China}
\affiliation{Beijing Key Laboratory for Magneto-Photoelectrical Composite and Interface Science, School of Mathematics and Physics, University of Science and Technology Beijing, Beijing 100083, China}

\begin{abstract}
The intrinsic coupling among electrical conductivity ($\sigma$), Seebeck coefficient ($S$), and lattice thermal conductivity ($\kappa_{\mathrm{L}}$) imposes a fundamental limit on the dimensionless figure of merit $ZT$ in thermoelectric (TE) materials. Increasing band degeneracy can effectively balance $\sigma$ and $S$, enabling a high power factor (PF, $S^{2}\sigma$). However, compounds with intrinsically large band degeneracy are scarce. Here, we present an unconventional strategy to realize elevated band degeneracy in zigzag-chain Na$_2$Au$X$ ($X$ = P, As, Sb, Bi) compounds by harnessing strong intra- and interchain orbital coupling. Pronounced hybridization between Au-$d_{z^{2}}$ and $X$-$p_{z}$ orbitals along the Au--$X$ zigzag chains, together with unexpectedly strong interchain $X$-$p_{x}/p_{y}$ coupling, produces a highly dispersive, multivalley valence band structure that supports an exceptional PF. Concurrently, the intrinsically weak interchain interactions arising from the quasi-one-dimensional framework, together with the weakened Au--$X$ and Au--Au bonds within the chains due to filling of $p$-$d^{*}$ antibonding states, result in an ultralow $\kappa_{\mathrm{L}}$. First-principles calculations combined with Boltzmann transport theory predict that $p$-type Na$_2$AuBi achieves a PF of $63.9\,\mu\mathrm{W}\,\mathrm{cm}^{-1}\,\mathrm{K}^{-2}$, an ultralow $\kappa_{\mathrm{L}}$ of $0.49\,\mathrm{W}\,\mathrm{m}^{-1}\,\mathrm{K}^{-1}$, and a maximum $ZT$ of $4.7$ along the zigzag-chain direction at $800\,\mathrm{K}$. This work establishes a new design paradigm for high-efficiency TE materials by exploiting substantial orbital overlap in structurally weakly bonded, quasi-one-dimensional systems, opening promising avenues for the discovery and engineering of next-generation high-performance TE materials.
\end{abstract}

\maketitle

\section{Introduction}\label{Introduction}
	Thermoelectric (TE) technology enables the direct conversion between heat and electricity, characterized by its simple structure, high reliability, and environmental friendliness. Owing to these advantages, TE technology finds broad applications in areas such as waste heat recovery~\cite{liang2025stable}, temperature-gradient power generation~\cite{hu2025all}, and microelectronics cooling~\cite{chen2022thermoelectric}, and is regarded as a promising solution in the field of renewable energy. Nevertheless, its large-scale deployment remains hindered by the inherently low energy-to-electricity conversion efficiency. The efficiency of TE materials is quantified by the dimensionless figure of merit $ZT = \frac{S^2 \sigma T}{\kappa_{\mathrm{L}} + \kappa_{\mathrm{e}}}$, where $T$ denotes the absolute temperature and $\kappa_{\mathrm{e}}$ represents the electronic thermal conductivity. According to the Wiedemann–Franz law~\cite{solidstatephysics}, $\kappa_{\mathrm{e}}$ is proportional to the electrical conductivity $\sigma$. Therefore, enhancing the power factor ($PF = S^2\sigma$) while reducing the lattice thermal conductivity $\kappa_{\mathrm{L}}$ is crucial for achieving high $ZT$ values~\cite{nolas1999skutterudites}. However, the intrinsic coupling between $S$ and $\sigma$ poses a significant challenge for improving TE performance. To date, many strategies have been developed to enhance TE efficiency. These include increasing the PF by optimizing carrier concentration~\cite{zhang2018deep,pei2014high,liu2025realizing}, band structure engineering~\cite{zhang2021band,pei2012band,guo2023enhanced,liu2012convergence}, and introducing resonant energy levels~\cite{zheng2022synergistically}. On the other hand, suppression of $\kappa_{\mathrm{L}}$ has been achieved through the incorporation of heavy-mass elements, introduction of defects and disorder, fabrication of nanostructures, enhancement of phonon scattering~\cite{B916400F,https://doi.org/10.1002/idm2.12134,ren2016contribution,zhao2017defect}, and weakening of chemical bonds~\cite{https://doi.org/10.1002/adfm.202108532,https://doi.org/10.1002/advs.202417292}.

	A high $S$ typically requires a large density-of-states effective mass ($m_{\mathrm{d}}^*$). The value of $m_{\mathrm{d}}^*$ can be increased either through elevated band degeneracy ($N_{\mathrm{v}}$) or a large band effective mass ($m_{\mathrm{b}}^*$). For a single parabolic band, $m_{\mathrm{d}}^* = N_{\mathrm{v}}^{2/3} m_{\mathrm{b}}^*$. However, an increased $m_{\mathrm{b}}^*$ generally leads to reduced carrier mobility ($\mu$), following the proportionality $\mu \propto (m_{\mathrm{b}}^*)^{-3/2} (m_{\mathrm{I}}^*)^{-1}$, where $m_{\mathrm{I}}^*$ is the inertial mass of carriers along the conduction direction~\cite{fu2014high}. Consequently, increasing $N_{\mathrm{v}}$ is an effective approach to simultaneously optimize $S$ and $\sigma$, thereby achieving a high power factor (PF)~\cite{2011Convergence,ti2022thermoelectric,he2019designing,xiong2025forbidden}. The $N_{\mathrm{v}}$ can be enhanced either when multiple bands become energetically degenerate at the Fermi level ($N_{\mathrm{vo}}$) or when multiple carrier pockets in the Brillouin zone exhibit degeneracy ($N_{\mathrm{vk}}$)~\cite{xiong2025forbidden}. For example, in PbTe, the convergence of the second maximum of the valence band (located at the midpoint of the $\Sigma$ line with $N_{\mathrm{vk}} = 12$) with the valence band maximum (VBM) at the L point ($N_{\mathrm{vk}} = 4$), achieved through alloying with PbSe, increases the $ZT$ from 0.8 to 1.8~\cite{2011Convergence}. Strategies for enhancing $N_{\mathrm{v}}$ include the use of lone-pair-electron cations~\cite{https://doi.org/10.1002/anie.201508381,https://doi.org/10.1002/adfm.202108532} and symmetry-forbidden $p$--$d$ coupling~\cite{xiong2025forbidden}, as evidenced in diverse material systems such as PbTe~\cite{PhysRevB.55.13605}, Li$_2$TlBi~\cite{he2019designing}, Bi$M$SeO ($M$ = Cu and Ag)~\cite{C4EE00997E,doi:10.1021/jacs.1c10284}, TlCu$X$ ($X$ = S and Se)~\cite{https://doi.org/10.1002/adma.202104908,doi:10.1021/jacs.4c16394}, and Pt$X_2$ ($X$ = P and As)~\cite{xiong2025forbidden}.
	
	While optimizing $N_{\mathrm{v}}$ can effectively balance $S$ and $\sigma$, achieving a high $ZT$ also requires lowering $\kappa_{\mathrm{L}}$, which introduces an additional challenge. In many solids, $\kappa_{\mathrm{L}}$ and $\sigma$ are intrinsically coupled: materials with low $\kappa_{\mathrm{L}}$ often possess low $\sigma$, whereas compounds with high $\sigma$ usually exhibit high $\kappa_{\mathrm{L}}$. This correlation, which hinders independent optimization, originates from the proportionality between $\sigma$ and the bulk modulus in the context of acoustic deformation potential (ADP) scattering~\cite{PhysRev.80.72}. A high bulk modulus reflects strong chemical bonding interactions, leading to elevated $\kappa_{\mathrm{L}}$~\cite{https://doi.org/10.1002/anie.201508381}, whereas weaker bonding generally corresponds to lower $\kappa_{\mathrm{L}}$. Layered and quasi-one-dimensional (1D) chain compounds present unique opportunities to circumvent this coupling. Due to extended bond lengths in directions perpendicular to the layer stacking or chain axis, these systems exhibit weaker orbital interactions in those orientations. For example, in layered compounds with van der Waals (vdW) gaps, negligible orbital overlap between adjacent layers results in very low $\sigma$ and $\kappa_{\mathrm{L}}$ along the stacking direction, as seen in BaTiS$_3$~\cite{PhysRevX.15.011066} and MoS$_2$~\cite{kim2021extremely,10.1063/1.4904513}. However, if the orbital coupling and bond lengths across the stacking or chain directions are carefully balanced, it is possible to decouple $\sigma$ from $\kappa_{\mathrm{L}}$, enabling the realization of high PF together with low $\kappa_{\mathrm{L}}$ in the same crystallographic orientation.
	
	In this study, we demonstrate that strong intra- and interchain orbital coupling in zigzag-chain compounds can simultaneously achieve high $N_{\mathrm{v}}$ and low $\kappa_{\mathrm{L}}$. First-principles calculations on Na$_2$Au$X$ ($X$ = P, As, Sb, and Bi) reveal that the multiband and highly dispersive valence-band edges primarily originate from pronounced orbital interactions between the $p_x$ and $p_y$ orbitals of $X$ in two adjacent $X$--Au--$X$ chains, as well as intrachain coupling between Au-$d_{z^2}$ and $X$-$p_z$ orbitals. In parallel, relatively weak chemical bonding between zigzag chains (due to the quasi-1D structure) and reduced Au--$X$ and Au--Au interactions, caused by the filling of $d$--$p^*$ antibonding states, lead to intrinsically low $\kappa_{\mathrm{L}}$. As a result, this family of compounds exhibits exceptionally high $ZT$ values, with a maximum $ZT$ of 4.7 observed in $p$-type Na$_2$AuBi at 800\,K. This finding expands our design principles for thermoelectric materials by demonstrating an effective pathway to decouple $\sigma$, $S$, and $\kappa_{\mathrm{L}}$, and opens new directions for achieving high-performance TE systems.

	\begin{figure*}[tph!]
	\includegraphics[width=1.0\linewidth]{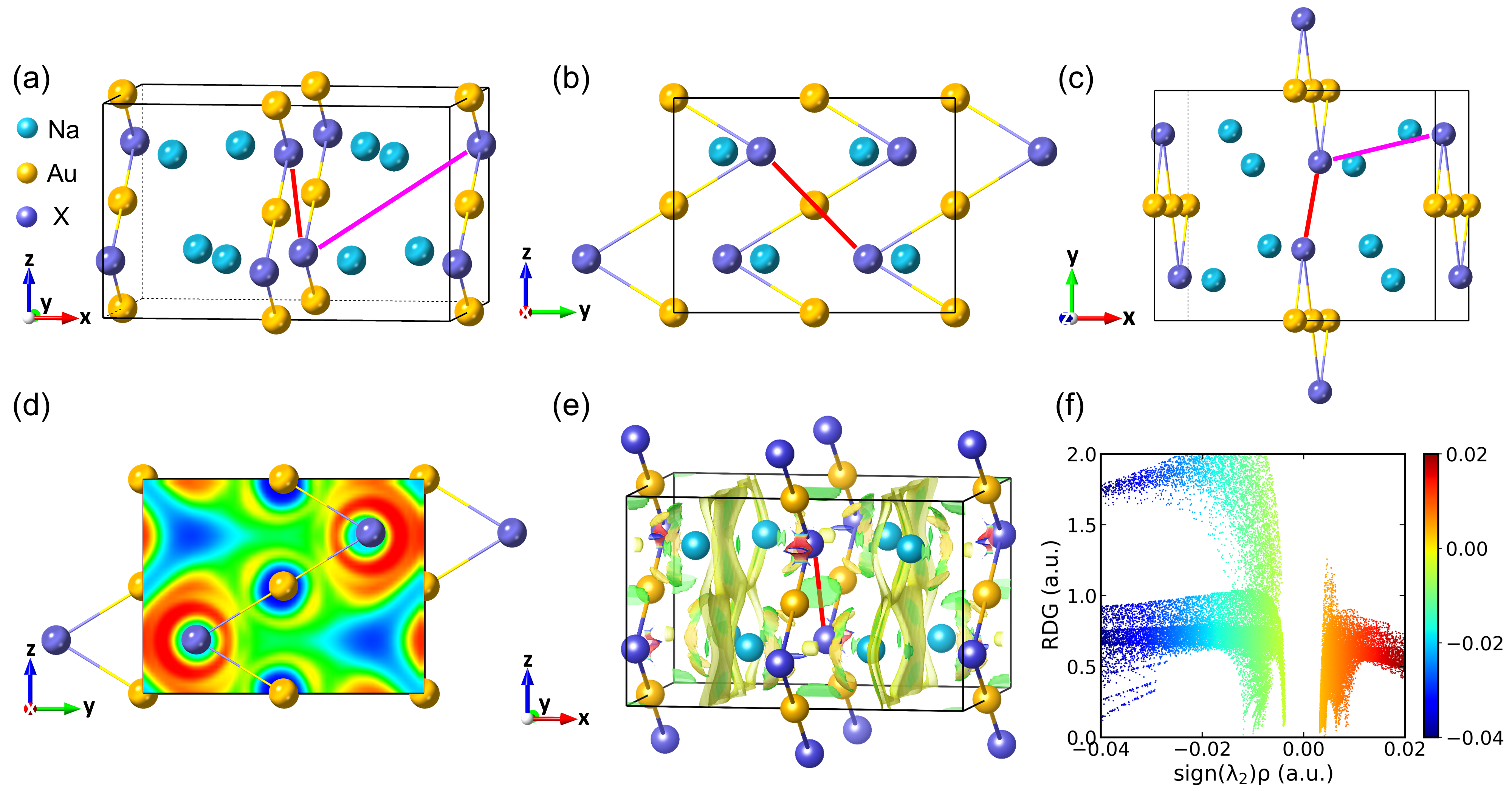}
	\caption{(a) The crystal structure of Na$_2$Au$X$ ($X$ = P, As, Sb, and Bi). (b) Top view of the Au-X zigzag chain along the $x$ direction within $y$-$z$ plane. (c) Side view of the Au-X zigzag chain along $z$ axis. The yellow, blue, and cyan balls represent Au, $X$ and Na atoms, respectively. The red and pink lines indicate the shortest distances between two Bi atoms that are from two Au-$X$ zigzag chains within a plan containing Au-$X$ zigzag chain and across the plan, respectively. (d) Electron localization function (ELF) of Na$_2$AuBi. (e) The RDG isosurface corresponds to RDG = 0.2 a.u., which is colored on a BGR scale of -0.04 $<\mathrm{sign(\lambda_2)\rho}<$ 0.02 a.u.. (f) RDG as a function of $\mathrm{sign(\lambda_2)\rho}$ for the atomic interactions in Na$_2$AuBi.}
	\label{structure}
    \end{figure*}

	\section*{Results and Discussion}\label{results_discussion}
    In these compounds, Na$_2$Au$X$ ($X$ = As, Sb, and Bi) were experimentally synthesized~\cite{mues1980na2auas,https://doi.org/10.1002/zaac.200900417} and Na$_2$AuP is predicted to be thermodynamically stable in the same structure by the open quantum materials database (OQMD)~\cite{kirklin2015open}. As shown in Figure~\ref{structure}, they crystallize in the orthorhombic structure with space group $Cmcm$ (No. 63), with Na, Au, and $X$ atoms occupying the Wyckoff positions of $8g$, $4b$, $4c$, which have site symmetry of $..m$, $2/m..$, $m2m$, respectively. This structure is characterized by a 1D infinite anion unit $_{\infty}$[AuX]$\mathrm{^{2-}}$ forming a Au-$X$ zigzag chains along the $z$-axis direction, which also can be viewed as that two Au atoms are linearly bonded to each other and are sharply curved at the $X$ atom. Note the zigzag chain also appears in alkali metal gold chalcogenides (RbAuS, CsAuS, RbAuSe) and gold halogenides (AuCl, AuI) compounds~\cite{bronger1992synthese,strahle1974kristalldaten,jagodzinski1959kristallstruktur}. The difference is that these chains are juxtaposed in the $yz$ plane to form $_{\infty}$[AuX]$\mathrm{^{2-}}$ sheets, which are separated by Na or other alkali metal layers. Our calculated lattice parameters together with the experimental lattice parameters for all compounds are tabulated in Table~\textcolor{red}{S1}. The good agreement between the calculated and experimental structural parameters indicates the reliability of our computational method.

	\begin{figure}[tph!]
		\includegraphics[width=1.0\linewidth]{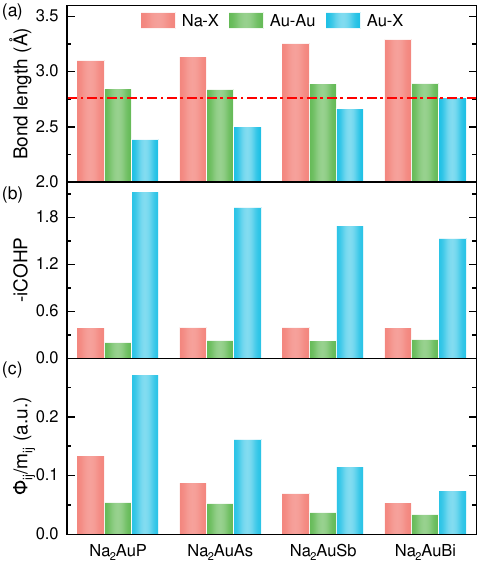}
		\caption{(a) The Na-$X$, Au-Au, and Au-$X$ bond lengths of Na$_2$Au$X$ compounds. The red dashed line indicates the Au-Au bond length of Na$_2$Au~\cite{havinga1972compounds}. (b) -iCOHP and (c) the ratio of interatomic force constant ($\Phi_{ij}$) to average mass ($m_{ij}$) for Na$_2$Au$X$ ($X$ = P, As, Sb, and Bi).}
		\label{cohp}
	\end{figure}

    \begin{figure*}[tph!]
	\includegraphics[width=1.0\linewidth]{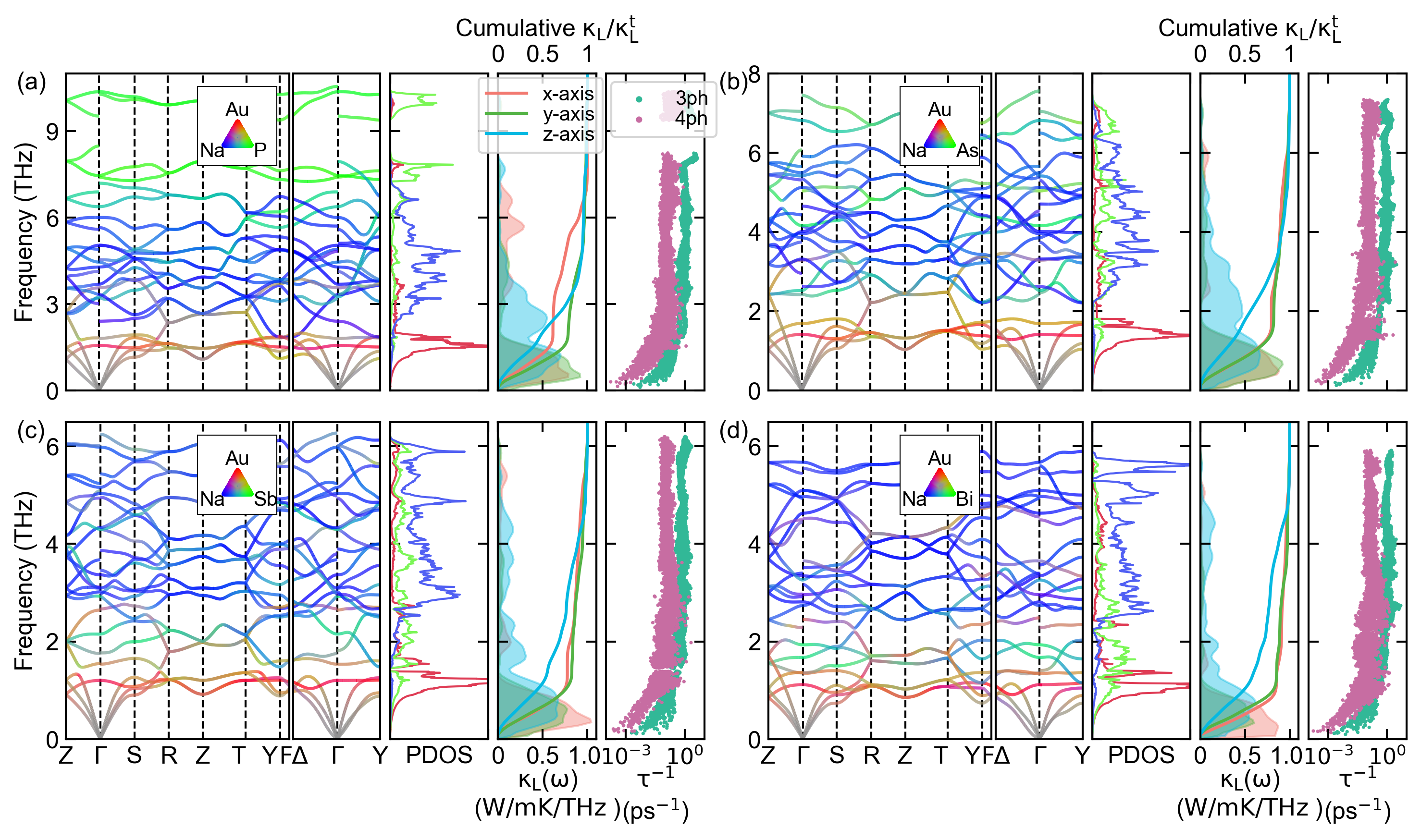}
	\caption{The phonon dispersion, phonon density of states, lattice thermal conductivity spectrum $\kappa_\mathrm{L}$($\omega$), the ratio of cumulative $\kappa_\mathrm{L}$ to total lattice thermal conductivity $\kappa_\mathrm{L}^{\mathrm{t}}$ and the 3ph and 4ph scattering rates ($\tau^{-1}$) for (a) Na$_2$AuP (b) Na$_2$AuAs (c) Na$_2$AuSb (d) Na$_2$AuBi.}
	\label{phonon}
    \end{figure*}

    \begin{figure*}[tph!]
    \includegraphics[width=1.0\linewidth]{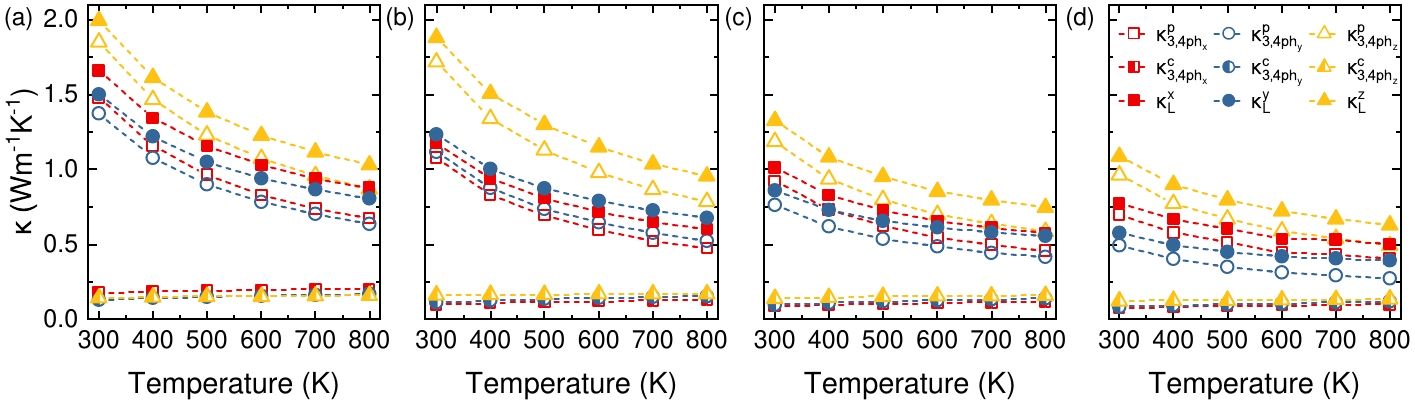}
    \caption{The calculated $\mathrm{\kappa_L}$ of Na$_2$AuX as a function of temperature in different directions, $\mathrm{\kappa^p}$, $\mathrm{\kappa^c}$ and $\mathrm{\kappa_L}$ are phonon population's contribution, additional coherence's contribution and total lattice thermal conductivity. (a), (b), (c), and (d) are Na$_2$AuP, Na$_2$AuAs, Na$_2$AuSb, and Na$_2$AuBi, respectively.}
    \label{kappa}
    \end{figure*}
    
	Figure~\ref{cohp} shows the bond lengths, integrated crystal orbital Hamiltonian population (iCOHP) and ratio of interatomic force constant to average mass ($\Phi_{ij}/m_{ij}$) for Na$_2$Au$X$. In these compounds, the bond length between Au and $X$ is within a distance ranging from 2.39 to 2.76 \AA, which closely matches the sum of the covalent radii (Au: 1.24 \AA, P: 1.11 \AA, As: 1.21 \AA, Sb: 1.40 \AA, Bi: 1.51 \AA)~\cite{pyykko2009molecular}, which indicates that Au-$X$ bond is more covalent, consisting with the plot of electron localization function (see Figure~\ref{structure}(d)). The non-covalent interaction (NCI) analysis~\cite{contreras2011nciplot}, based on the electron density and its derivatives, provides an effective tool for visualizing and characterizing weak interactions in molecular and crystalline systems. This method utilizes the reduced density gradient (RDG), plotted as a function of the quantity sign($\lambda_2$)$\rho$, where $\rho$ is the electron density and sign($\lambda_2$) denotes the sign of the second eigenvalue of the electron density Hessian matrix. In this representation, negative values of sign($\lambda_2$)$\rho$ indicate attractive interactions, while positive values correspond to repulsive interactions. Characteristic spikes at low RDG values in regions of low electron density signify non-covalent contact points, with the magnitude of $\rho$ at these spikes offering a qualitative measure of interaction strength. This approach enables the intuitive identification and analysis of weak intermolecular forces in complex systems. As shown in Figure~\ref{structure}(e), the 3D RDG isosurface, mapped on a blue–green–red color scale, reveals distinct NCI. The green color indicates that the attractive interaction exists between the Bi atoms of the two Au-Bi atomic chains within the $yz$ plane. Between two Au-$X$ chains along the $x$ direction, there is an obvious isosurface parallel to the chain direction, which comes from the widely distributed van der Waals (vDW) interaction between the layers. While the red and blue isosurfaces are mainly distributed between Au-Bi and Au-Au, indicating stronger interactions within the chains. To quantitatively understand the interlayer interactions, the dependence of RDG on sign($\lambda_2$)$\rho$ is shown in Figure ~\ref{structure}(f). When sign($\lambda_2$)$\rho$ is approaching zero (interlayer interaction, low $\rho$), RDG is green, indicting the vDW interaction.

	 We also calculate the ionic character of bonds (ICB)~\cite{doi:10.1021/j150495a016} of Au-$X$ bond in these compounds and find the ionicity of Au-$X$ bond is at most 6.5 \%, suggesting a strong polar covalent character, due to high electronegativity of Au. In addition, previous studies have shown that when a bridging ligand $X$ is present between two cations $M$ with a $d^{10}$ configuration, the bridging ligand significantly reduces the interaction of the $M$-$M$ bond compared to that without the bridging ligand $X$~\cite{cui1990bonding}. Therefore, the Au-Au bond length in these compounds is slightly greater than that observed in Na$_2$Au (2.76 \AA)~\cite{havinga1972compounds}, which contains only a single Au atoms chain. The -iCOHP can be used to characterize the strength of chemical bonds. Among these compounds, the distance between two Au atoms within the zigzag chain is shorter than that between Na and $X$, but the -iCOHP and $\Phi_{ij}/m_{ij}$ of Na-$X$ bond is larger than that of Au-Au, due to the present of bridge ligand. The Au-$X$ bond within the Au-$X$ zigzag chain has the shortest bond length in each compound, and therefore has the largest -iCOHP and $\Phi_{ij}/m_{ij}$, indicating Au-$X$ has the strongest bonding interaction. Bond length, -iCOHP, and $\Phi_{ij}/m_{ij}$ consistently show that the bonding strengths of all these chemical bonds continuously decline from P to Bi, indicating a decreasing speed of sound ($\nu_{\mathrm{g}}$) and therefore a decreasing $\kappa_{\mathrm{L}}$ with the mass of $X$ increasing. The shortest distances between two $X$ atoms within two adjacent Au-$X$ zigzag atomic chains (in the $yz$ plane) is in the range of 4.09-4.16 \AA\, (varying with $X$), while the shortest distance between two $X$ atoms within two adjacent layers (along the $x$-direction) is from 5.20 to 5.62 \AA, as indicated by the red and pink lines in Figure~\ref{structure}, respectively.

    Phonon dispersion and phonon density of states (PhDOS) of Na$_2$Au$X$ ($X$ = P, As, Sb, and Bi) compounds calculated at 300 K are illustrated in Figure~\ref{phonon}, while the phonon spectra calculated at 0 K are shown in Figure~\textcolor{red}{S1}. The absence of imaginary phonon modes at both 0 K and 300 K indicates the structure is in the local minimal of the potential energy surface and there is no phase transition within this temperature range. Since the primitive cell of Na$_2$Au$X$ has 8 atoms, there are 3 acoustic and 21 optical phonon branches. As $X$ varies from P to Bi, the gradual decrease in electronegativity and increase in ionic radius of $X$ lead to weaker and weaker bonding interactions, and a corresponding reduction in the highest optical phonon frequencies. Overall, both the lowest and highest frequency of optical phonons of these compounds at the $\Gamma$ point are relatively low, indicating weak bonding interactions and low $\nu_{\mathrm{g}}$ in these compounds. Moreover, there is an obvious anisotropy in the phonon dispersion, the longitude acoustic (LA) mode along the $\Gamma$-Z direction (the $z$ axis in real space, see Figure \textcolor{red}{S5}) exhibits significantly higher frequencies compared to that along the $\Gamma$-$\Delta$ (the $y$ axis in real space) and $\Gamma$-Y (the $x$ axis in real space) directions, suggesting a higher $\nu_{\mathrm{g}}$ and therefore potentially higher $\kappa_\mathrm{L}$ along the $z$ axis. This is consistent with the conclusion above based on chemical bond that the Au-$X$ bond in the zigzag chains has the strongest chemical bonding interaction. Along the $\Gamma$-$\Delta$ and $\Gamma$-Y directions, pronounced avoided crossing between acoustic and optical modes are observed in all these compounds, which significantly reduce the highest frequency of the LA phonon mode at zone boundary and indicate strong acoustic-optical phonon interaction along the direction perpendicular to the Au-$X$ atomic chain. As shown in Figure~\textcolor{red}{S2}, though Na atom exhibits larger atomic displacement parameters (ADPs) at the $y$ and $z$ directions, due to its small atomic mass and weaker Na-$X$ bonds, the ADPs of Au atoms along the $x$ direction is comparable or even larger than that of Na, indicating relatively weaker chemical bonding interactions with Au along the out-of-the-plane (the $x$ axis) direction. This is because the stronger constraints within Au-$X$ zigzag chain and among the chain within the zigzag chain plane, Au atoms only can vibrate outside of the Au-$X$ zigzag chain plane, as evidenced by the Au-dominated flat phonon bands along the $\Gamma$-Y direction. As shown in Figure~\ref{phonon}, sharp peaks in the low-frequency region (1 $\sim$ 2 THz) of PhDOS are mainly attributed by Au atoms.

 	\begin{figure*}[tph!]
 	\includegraphics[width=1.0\linewidth]{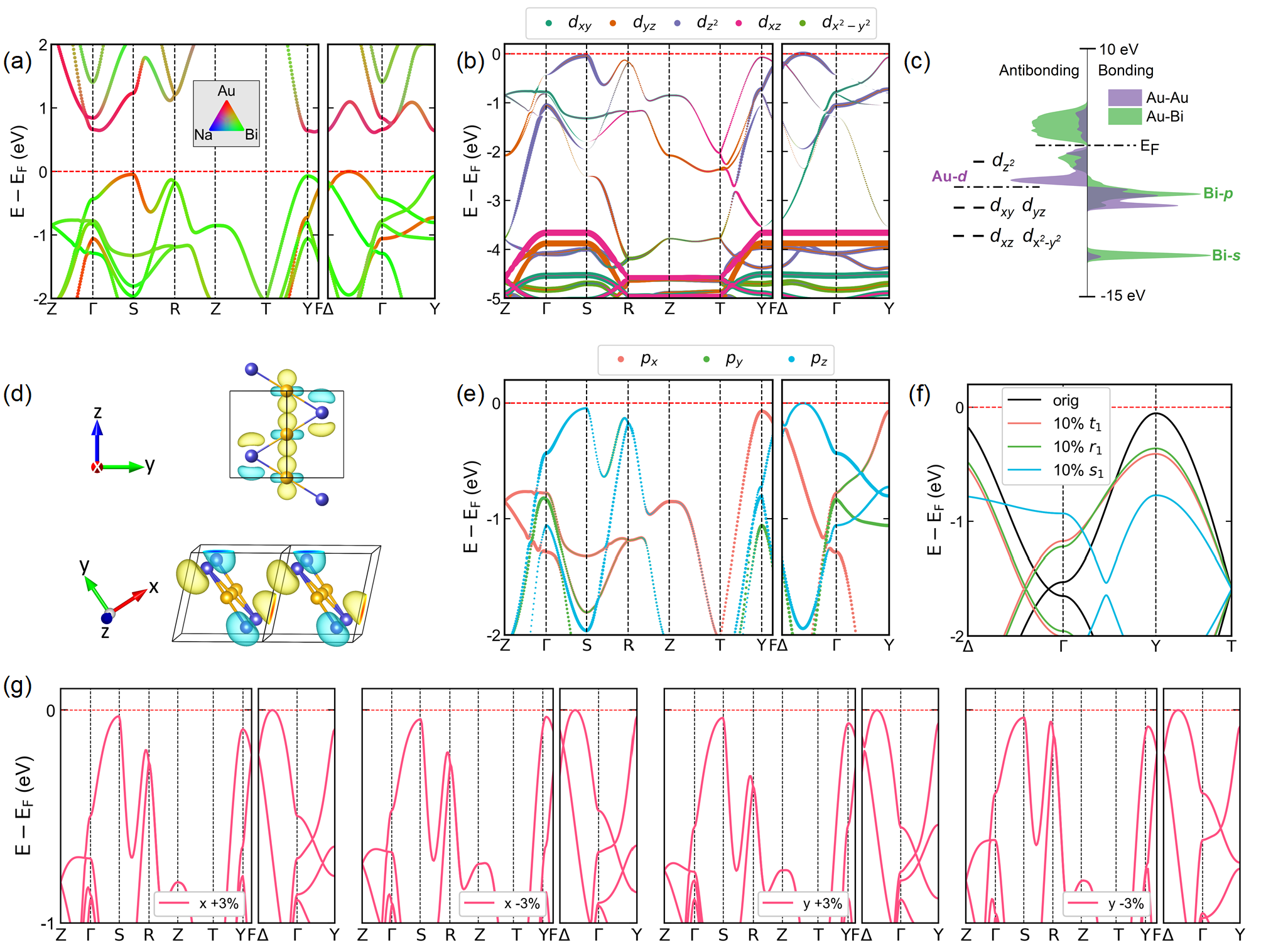}
 	\caption{(a) The element projected band structure of Na$_2$AuBi. (b) The $d$-orbital projected band structure of Au atoms in Na$_2$AuBi. (c) The -COHP and schematic diagram of linear chain crystal field. (d) The maximally-localized Wannier functions of Au-$d_{z^2}$ with Bi-$p_z$ (interchain) and Bi-$p_x$ (intrachain). (e) The $p$-orbital projected band structure of Bi atoms in Na$_2$AuBi. (f) The Tight binding band model under different parameter fitting. (g) The valence band structure of Na$_2$AuBi under different stresses.}

 	\label{wannier}
    \end{figure*}

The Boltzmann phonon transport equation was solved using a self-consistent iterative method. The $\kappa_\mathrm{L}$, incorporating corrections to the second-order force constants at finite temperatures, three- and four-phonon scattering processes, as well as the coherent contribution, is calculated over the temperature range of 300--800~K, as shown in Figure~\ref{kappa}. The decrease of $\kappa_\mathrm{L}$ with increasing temperature closely follows a $T^{-1}$ trend. Owing to the anisotropic nature of chemical bonding in different crystallographic directions, as discussed above, $\kappa_\mathrm{L}$ exhibits pronounced directional dependence. Consistent with this analysis, all compounds display the largest $\kappa_\mathrm{L}$ along the $z$-direction, corresponding to the Au--$X$ zigzag chain orientation. For Na$_2$AuAs, the minimum $\kappa_\mathrm{L}$ occurs along the $x$-direction, whereas in the other compounds it appears along the $y$-direction, although the difference between the $x$ and $y$ directions is negligible. Among the investigated systems, Na$_2$AuBi exhibits the lowest $\kappa_\mathrm{L}$ along all three directions and, consequently, the lowest average $\kappa_\mathrm{L}$ (0.81~W$\mathrm{\,m^{-1}K^{-1}}$) at room temperature. This value is significantly smaller than those of well-established thermoelectric materials, such as Mg$_3$Sb$_2$ (1.6~W$\mathrm{\,m^{-1}K^{-1}}$)~\cite{li2021demonstration}, PbTe (2.2~W$\mathrm{\,m^{-1}K^{-1}}$)~\cite{akhmedova2009effect}, and SrAgSb (1.9~W$\mathrm{\,m^{-1}K^{-1}}$)~\cite{zhang2020promising}.

The differential lattice thermal conductivity ($\kappa_\mathrm{L}(\omega)$) represents the contribution of phonons with frequency $\omega$ to the total lattice thermal conductivity ($\kappa_\mathrm{L}^{\mathrm{t}}$). The ratio of cumulative $\kappa_\mathrm{L}$ to $\kappa_\mathrm{L}^{\mathrm{t}}$ is presented in Figure~\ref{phonon}. The results indicate that heat transport along the $x$- and $y$-axes is predominantly governed by acoustic and low-frequency optical modes below 1.7~THz, which together contribute more than 70\% of $\kappa_\mathrm{L}^{\mathrm{t}}$. The remaining contribution arises from mid-frequency phonons, where Na atom vibrations are dominant. In Na$_2$AuP, these mid-frequency modes contribute relatively more to $\kappa_\mathrm{L}$ along the $x$-axis compared with the other compounds. In contrast, $\kappa_\mathrm{L}$ along the $z$-axis exhibits a steady increase across the 0--5~THz frequency range, where the dominant heat-carrying modes originate from both Au and Na atoms. This can be attributed to the strong chemical bonding along the $z$-direction, aligned with the Au--$X$ zigzag chains, which is responsible for the higher $\kappa_\mathrm{L}$ in that direction.

Within the framework of kinetic theory~\cite{tritt2005thermal}, $\kappa_\mathrm{L}$ is given by  
\begin{equation}
	\kappa_\mathrm{L} = \frac{1}{3} C_{\mathrm{v}} \nu_{\mathrm{g}}^2 \tau ,
\end{equation}
where $C_{\mathrm{v}}$ is the specific heat, $\nu_{\mathrm{g}}$ is the phonon group velocity, and $\tau$ is the phonon relaxation time. As shown in Figure~\textcolor{red}{S4}, the investigated compounds exhibit very similar $C_{\mathrm{v}}$ values. Therefore, $\kappa_\mathrm{L}$ is primarily determined by $\nu_{\mathrm{g}}$ and $\tau$. The reduction in electronegativity from P to Bi leads to weaker chemical bonding strength ($k$) and higher atomic mass ($M$), which collectively lower $\nu_{\mathrm{g}}$ according to $\nu_{\mathrm{g}} \propto \sqrt{k/M}$ (Table~S2). The calculated phonon--phonon scattering rates, $\tau^{-1}$, at 300~K are shown in Figure~\ref{phonon}. Higher $\tau^{-1}$ corresponds to shorter phonon lifetimes and, consequently, to significantly reduced $\kappa_\mathrm{L}$. Three-phonon (3ph) scattering dominates in all compounds, while strong four-phonon (4ph) scattering appears only in the low-frequency region ($\sim$1~THz) associated with flat phonon branches. These flat-band features correspond to two distinct Au peaks in the phonon density of states (PhDOS), as displayed in Figure~\ref{phonon}. Na$_2$AuBi exhibits the strongest phonon--phonon scattering, attributed to its large weighted phase space, which provides more available scattering channels (Figure~\textcolor{red}{S3}). The combined effect of shorter $\tau$ from enhanced scattering and lower $\nu_{\mathrm{g}}$ results in the exceptionally low $\kappa_\mathrm{L}$ observed in Na$_2$AuBi.

	\begin{figure*}[tph!]
	\includegraphics[width=1.0\linewidth]{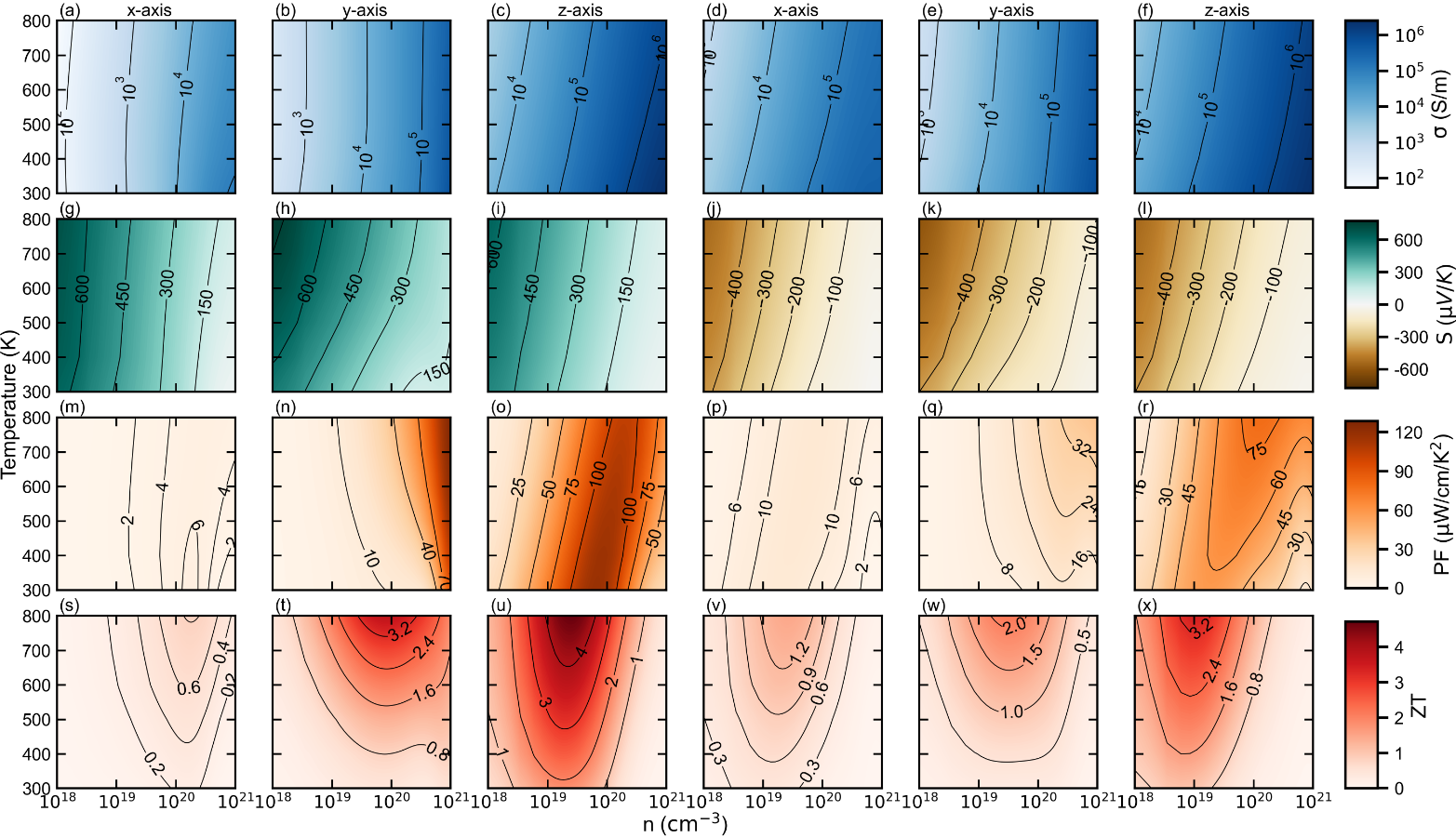}
	\caption{The contour plots of the electron transport properties of Na$_2$AuBi along the $x$, $y$, and $z$ directions as a function of carrier concentration and temperature. The conductivity $\sigma$ (a)-(f), Seebeck coefficient $S$ (g)-(l), power factor PF (m)-(r) and thermoelectric figure of merit ZT (s)-(x) for $p$-type (left three columns) and n-type (right three columns) doping, respectively.}
	\label{PF}
    \end{figure*}

    \begin{figure*}[tph!]
	\includegraphics[width=1.0\linewidth]{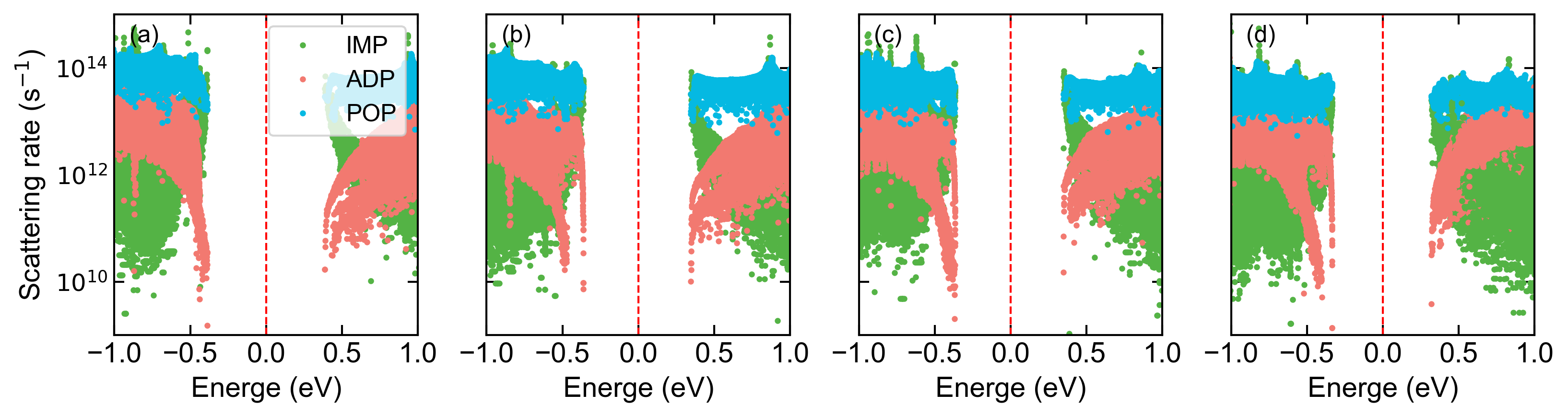}
	\caption{The ADP, IMP and POP scattering rates contribution near the conduction and valence band edges as a function of energy at room temperature for (a) Na$_2$AuP (b) Na$_2$AuAs (c) Na$_2$AuSb (d) Na$_2$AuBi.}
	\label{sc}
    \end{figure*}

    Since the electronic structures of Na$_2$Au$X$ ($X$ = P, As, Sb, and Bi) are very similar, only the band structure of Na$_2$AuBi with orbital contributions indicated by color coding is shown in Figure~\ref{wannier}(a), while the remaining results are presented in Figures~\textcolor{red}{S6--S8}. Accurate determination of the band gap and band dispersions near the Fermi level is essential for evaluating the thermoelectric (TE) performance of these materials. It is well known that semilocal exchange--correlation functionals, such as PBE, tend to underestimate the band gap. Therefore, we employed the screened hybrid functional HSE06~\cite{10.1063/1.1564060} to compute the band gaps of these compounds, with the results summarized in Table~\textcolor{red}{S4}. All four compounds are identified as indirect band gap semiconductors. For Na$_2$AuP, the valence band maximum (VBM) is located at the Y point, whereas for the remaining compounds, the VBM lies along the $\Gamma$--$\Delta$ line. The conduction band minimum (CBM) for Na$_2$AuP, Na$_2$AuAs, and Na$_2$AuSb is found at the $\Gamma$ point, while that of Na$_2$AuBi is positioned along the $\Gamma$--$\Delta$ line and is nearly degenerate with a state located on the Y--F path.

    For all compounds, the valence band edge is dominated by Au-$d$ and $X$-$p$ states, reflecting strong $d$--$p$ orbital hybridization. Remarkably, both valence and conduction bands exhibit large dispersions at the band edges, even along directions perpendicular to the Au--$X$ zigzag chains, such as $\Gamma$--Y and $\Gamma$--$\Delta$. Since the Au--$X$ chain lies in the $yz$ plane, the Au--$X$ interactions within a chain and between chains in the $yz$ plane are stronger than those along the $x$ direction. Consequently, the energy at the Y point is slightly lower than that in the middle of the $\Gamma$--$\Delta$ line. Although bands along $\Gamma$--Z (parallel to the Au--$X$ chains) lie far below the Fermi energy, certain bands along symmetry-equivalent directions such as T--Y and R--S are close to the Fermi level, confirming the multiband nature of these systems. As shown in Figures~\ref{wannier}(b) and \ref{wannier}(e), the highly dispersive valence bands along $\Gamma$--S and $\Gamma$--$\Delta$ originate from strong coupling between Au-$d_{z^2}$ and Bi-$p_z$ orbitals along the Au--Bi zigzag chain. Figure~\ref{wannier}(c) reveals that both Au--Bi and Au--Au bonds exhibit antibonding states near the Fermi level, whereas the remaining $d$ orbitals are located below $-3.8$~eV, consistent with crystal-field splitting in a quasi-linear chain geometry. The maximally localized Wannier functions (MLWFs) confirm significant overlap between Au-$d_{z^2}$ and Bi-$p_z$ orbitals along the $z$-axis (Figure~\ref{wannier}(d)). Furthermore, strong overlap is observed between $p_x$ orbitals of Bi atoms in adjacent layers. This $p_x$--$p_x$ coupling leads to large dispersions ($\sim 2$~eV) along the T--Y and $\Gamma$--Y directions.

    To further elucidate the influence of interlayer bond distances perpendicular to the Au--$X$ chain, we applied uniaxial strain along the $x$ and $y$ directions and calculated the band structures of Na$_2$AuBi. As shown in Figure~\ref{wannier}(g), a 3\% strain significantly modifies the band dispersion, particularly the energy at the Y point for $x$-axis strain and at the R point for $y$-axis strain. This is because the Bi--Bi distances, both within adjacent layers of a single Au--$X$ zigzag chain (pink lines in Figure~\ref{structure}) and between neighboring zigzag chains (red lines in Figure~\ref{structure}), are highly sensitive to the lattice constants along $x$ and $y$ directions, respectively. Therefore, the multiband and strongly dispersive character of Na$_2$Au$X$ arises from a combination of strong Au-$d_{z^2}$--Bi-$p_z$ coupling within the zigzag chains and unexpectedly strong Bi-$p_x$--Bi-$p_x$ coupling perpendicular to the chain direction.

    To investigate the origin of this coupling, we constructed a tight-binding (TB) model considering only the Bi-$p_x$ orbital. Hopping interactions between first-, second-, and third-nearest neighbors were included (red lines, Bi--Au--Bi, and pink lines in Figure~\ref{structure}), with parameters $t_1 = -0.197$, $r_1 = -0.171$, and $s_1 = 0.199$. The effective Hamiltonian of the model is given by:
    \begin{equation}
	H(\mathbf{k}) =
	\begin{pmatrix}
		e_1 & 2 \cos \left( \frac{k_{z}}{2} \right) f(\mathbf{k}) \\
		2 \cos \left( \frac{k_{z}}{2} \right) f^{*}(\mathbf{k}) & e_1
	\end{pmatrix}
    \end{equation}
    with
    \begin{equation}
	\begin{split}
		f(\mathbf{k}) = & \ r_1 \, e^{-i\left(\frac{3 k_{x}}{5}+\frac{3 k_{y}}{5}\right)} 
		+ s_1 \, e^{i\left(\frac{2 k_{x}}{5}-\frac{3 k_{y}}{5}\right)} \\
		& + s_1 \, e^{i\left(-\frac{3 k_{x}}{5}+\frac{2 k_{y}}{5}\right)} 
		+ t_1 \, e^{i\left(\frac{2 k_{x}}{5}+\frac{2 k_{y}}{5}\right)} .
	\end{split}
    \end{equation}

    The TB-derived band structure is shown in Figure~\ref{wannier}(f). The model successfully reproduces the dispersions along $\Gamma$--Y and T--Y, confirming the dominant role of Bi-$p_x$ orbitals in the valence band. By selectively reducing one hopping parameter at a time, we find that all three parameters contribute to the large dispersion along $\Gamma$--Y--T. However, $s_1$, corresponding to interactions between two Au--Bi chains perpendicular to the $x$ direction (pink lines in Figure~\ref{structure}), has the strongest effect. This finding is consistent with the nearly parallel MLWFs of the two Au--Bi chains observed in Figure~\ref{wannier}.

	The $\sigma$, $S$, and PF of these compounds are calculated over a carrier concentration ($n$) range from $10^{18}$ to $10^{21}~\mathrm{cm^{-3}}$ and temperatures between 300 and 800~K, for both $n$-type and $p$-type doping. Since the electron transport properties are comparable among the compounds, only the transport data for Na$_2$AuBi are shown in Figure~\ref{PF}, while the results for the remaining systems are provided in Figures~\textcolor{red}{S9--S11} of the supplementary information. For both electron and hole doping, $\sigma$ exhibits pronounced anisotropy, with the highest values observed along the $z$-axis. This is primarily attributed to the smallest hole ($m^*_{\mathrm{h}}$) and electron ($m^*_{\mathrm{e}}$) effective masses along this direction, as listed in Table~\textcolor{red}{S4}. The Seebeck coefficient $S$ of Na$_2$AuBi also shows significant anisotropy. The maximum $S$ (773.1~$\mathrm{\mu VK^{-1}}$) occurs along the $y$-axis at a hole concentration $n_{\mathrm{h}} \approx 10^{18}~\mathrm{cm^{-3}}$ and $T = 800$~K. This enhancement arises because the valence band edges lie predominantly along the $\Gamma$--$\Delta$ line, which exhibits a valley degeneracy of 2. In contrast, the minimum $S$ (628.8~$\mathrm{\mu VK^{-1}}$) is recorded along the $z$-axis. Additionally, the highest valence bands at Y and S are close in energy to the VBM along $\Gamma$--$\Delta$, further increasing the band degeneracy. Due to the combination of high $\sigma$ and relatively large $S$ along the $z$-axis, the PF in this direction is superior to those along the $x$- and $y$-axes for both $p$-type and $n$-type doping. For example, the maximum PF of Na$_2$AuBi along the $z$ axis occurs at $T = 500$~K and $n_{\mathrm{p}} \approx 1.1 \times 10^{20}~\mathrm{cm^{-3}}$, reaching 113.7~$\mathrm{\mu W\,cm^{-1}K^{-2}}$, which is substantially higher than the values along the $x$-axis (6.4~$\mathrm{\mu W\,cm^{-1}K^{-2}}$) and $y$-axis (94.2~$\mathrm{\mu W\,cm^{-1}K^{-2}}$). For Na$_2$AuP, Na$_2$AuAs, and Na$_2$AuSb, the maximum PF also occurs along the $z$ axis, with respective values of 66.9~$\mathrm{\mu W\,cm^{-1}K^{-2}}$ (at 600~K and $4.0\times10^{20}~\mathrm{cm^{-3}}$), 77.4~$\mathrm{\mu W\,cm^{-1}K^{-2}}$ (500~K and $1.4\times10^{20}~\mathrm{cm^{-3}}$), and 85.2~$\mathrm{\mu W\,cm^{-1}K^{-2}}$ (500~K and $1.6\times10^{20}~\mathrm{cm^{-3}}$). For $n$-type doping, the PF along the $z$ axis shows a non-monotonic trend: it initially decreases and then increases with rising $n_{\mathrm{e}}$. The highest $n$-type PF of 79.5~$\mathrm{\mu W\,cm^{-1}K^{-2}}$ is obtained at $T = 800$~K and $n_{\mathrm{e}} \approx 1.3\times10^{20}~\mathrm{cm^{-3}}$ along the $z$-axis.

	Figure~\ref{sc} presents the electron--phonon scattering rates as a function of energy near the valence and conduction band edges at room temperature. Among the scattering mechanisms considered, polar optical phonon (POP) scattering exerts a significantly stronger influence on electron transport than both acoustic deformation potential (ADP) and ionized impurity (IMP) scattering, with ADP scattering consistently playing a minor role. The ADP mechanism arises from elastic strain fields within the crystal, whereas POP scattering is strongly temperature-dependent, with its contribution increasing progressively at elevated temperatures. The intensity of POP scattering is proportional to $\frac{1}{\epsilon^{\infty}} - \frac{1}{\epsilon^{0}}$ and $\omega_{\mathrm{pop}}$~\cite{ganose2021efficient}, where $\epsilon^{\infty}$ and $\epsilon^{0}$ denote the high-frequency and static dielectric constants, respectively, and $\omega_{\mathrm{pop}}$ is the effective POP frequency. As shown in Table~\textcolor{red}{S3}, both $\frac{1}{\epsilon^{\infty}} - \frac{1}{\epsilon^{0}}$ and $\omega_{\mathrm{pop}}$ decrease steadily from Na$_2$AuP to Na$_2$AuBi. Consequently, the PF displays an inverse trend with respect to $\frac{1}{\epsilon^{\infty}} - \frac{1}{\epsilon^{0}}$. For all compounds, $\frac{1}{\epsilon^{\infty}} - \frac{1}{\epsilon^{0}}$ along the $z$-axis is slightly smaller than along the $x$- and $y$-axes, resulting in enhanced $\sigma$ and, in turn, a larger PF in the $z$-axis direction. As discussed above, a 3\% compressive strain along the $y$-axis can increase the band degeneracy of Na$_2$AuBi. To further explore this effect, we computed the PF of Na$_2$AuBi under a 3\% compression strain and found that it is enhanced by approximately 65\% compared with the unstrained case, as shown in Figure~\textcolor{red}{S12}.

	The calculated $ZT$ values of Na$_2$AuBi for different crystallographic orientations are shown in Figures~\ref{PF}(s)--(x), while those of the other compounds are provided in Figures~\textcolor{red}{S9--S11}. Owing to the pronounced anisotropy of $\kappa_{\mathrm{L}}$ and PF under both $n$-type and $p$-type doping, the $ZT$ values vary substantially with direction for the same carrier concentration and temperature. For Na$_2$AuBi, the highest $ZT$ values for $p$-type and $n$-type doping are 4.7 and 3.4, respectively, attained along the $z$-axis at 800~K with optimized carrier concentrations of $2\times10^{19}$ and $1\times10^{19}~\mathrm{cm^{-3}}$. Under $p$-type doping, the $x$- and $y$-axis $ZT$ values reach 0.8 and 3.7, respectively, at 800~K, while under $n$-type doping, they reach 1.4 and 2.1, respectively. A similar trend is observed for Na$_2$AuP, Na$_2$AuAs, and Na$_2$AuSb. Specifically, the maximum $ZT$ values under $p$-type doping along $z$ are 2.5, 2.9, and 3.6, respectively, at 800~K, whereas the corresponding $n$-type values are 1.6, 2.0, and 2.4, respectively. Across the series Na$_2$Au$X$ ($X$ = P, As, Sb, Bi), the highest $ZT$ increases monotonically from P to Bi, due to the synergistic effects of enhanced $\sigma$ and reduced $\kappa_{\mathrm{L}}$. Overall, these compounds demonstrate outstanding thermoelectric performance in both $p$-type and $n$-type regimes. Although $p$-type doping yields higher $ZT$ values than $n$-type doping, the optimal carrier concentration required for $n$-type is slightly lower, making it potentially more feasible experimentally. Notably, our calculated $ZT$ values for Na$_2$Au$X$ compounds significantly exceed that of the well-studied Zintl phase Mg$_3$Sb$_2$ (1.88), computed using identical methodologies~\cite{han2025weak} and in excellent agreement with the experimental value of 1.55~\cite{doi:10.1073/pnas.1711725114}. These results highlight Na$_2$Au$X$ compounds as promising candidates for high-efficiency thermoelectric applications.

    \section*{Conclusion}\label{conclusion}
    In summary, we have demonstrated that quasi-one-dimensional Na$_2$Au$X$ ($X$ = P, As, Sb, and Bi) compounds possess a rare combination of highly dispersive multiband electronic structures and ultralow lattice thermal conductivity. The strong intra-chain Au-$d$/$X$-$p$ hybridization within the Au--$X$ zigzag chain, together with pronounced inter-chain $p_x$ coupling, reduces the carrier effective mass and enhances valley degeneracy, enabling large power factors for both $p$- and $n$-type doping. Concurrently, the weak Au--$X$ and Au--Au bonding interactions along the zigzag chains, stemming from $p$-$d^*$ antibonding states occupation and reduced $d$-$d$ interaction, and minimal interchain coupling suppress heat transport to an exceptional degree. These cooperative effects drive $ZT$ values as high as 4.7 ($p$-type) and 3.4 ($n$-type) in Na$_2$AuBi at 800~K. Our results not only identify Na$_2$Au$X$ as promising thermoelectric candidates but also introduce a general design concept for high-efficiency thermoelectrics: leveraging weak chemical bonding in conjunction with strong orbital overlap to synergistically optimize charge and phonon transport. This strategy opens new avenues for the discovery and engineering of next-generation thermoelectric materials with high performance.


    \section{COMPUTATIONAL DETAILS}\label{experimental}
    The first-principles calculations were performed by using density functional theory (DFT), as implemented in the Vienna ab initio simulation package (VASP)\cite{vasp1,vasp2}. The projector augmented wave (PAW) pseudo potentials\cite{paw1,paw2}, plane wave basis set, and the PBEsol exchange-correlation functional was used for crystal structure relaxation and phonon related properties calculations\cite{pbesol1}. The plane-wave kinetic-energy cutoff is set at 520 eV. In structural optimization, the energy and force tolerances are 10$^{-8}$ eV and 10$^{-3}$ eV/\AA, respectively. The elastic constants and mechanical properties were obtained through finite differences methods. The dielectric constants have been calculated using density functional perturbation theory (DFPT). The nature of chemical bond analysis based on the Crystal Orbital Hamilton Population (COHP) is investigated using the LOBSTER code\cite{nelson2020lobster}. The transport effective mass were calculated using BoltzTraP2 according to $m^*(T,\epsilon)=\frac{e^2\tau}{\sigma(T,\epsilon)}n(T,\epsilon)$. The maximally-localized Wannier functions including the $d$-orbitals of Au atom and the $p$-orbitals of $X$ atom were constructed on a $k$-mesh of 9 $\times$ 9 $\times$ 9, using the Wannier90 code~\cite{mostofi2008wannier90}. Based on the results of the Wannier orbital projection analysis, a tight-binding model was constructed using the MagneticTB software~\cite{zhang2022magnetictb}. The reduced density gradient (RDG)~\cite{contreras2011nciplot} analysis was performed as implemented in the Quantum Espresso~\cite{giannozzi2017advanced}.

    Further, the second force constants were computed by using the Phonopy with 4 $\times$ 4 $\times$ 4 supercell and $\Gamma$ k-points mesh\cite{phonopy-phono3py}. The machine learning accelerated $ab$ $initio$ molecular dynamics simulation was run for 20000 steps with a step of 1 fs at 300 K, and the 20 random configurations were extracted at equal intervals during the energy convergence stage. Furthermore, accurate DFT calculations were performed for these configurations to obtain the training and cross-validation force and displacement datasets, and the third and fourth force constants are extracted by the compressive sensing lattice dynamics (CSLD) techniques\cite{zhou2014lattice}. The temperature-dependent anharmonic phonon was computed with the self-consistent phonon theory (SCPH)\cite{werthamer1970self} considering the quartic anharmonic phonon renormalization. The lattice thermal conductivity was calculated by solving the linearized Boltzmann transport equation using FourPhonon\cite{han2022fourphonon}, including three-phonon (3ph) and four-phonon (4ph) interactions.

    The electron-phonon scattering matrix and electron transport properties are evaluated using the $ab$ $initio$ scattering and transport (AMSET) code\cite{ganose2021efficient}. Specifically, scattering-mechanism-dependent transport properties are examined under the consideration of scattering from acoustic deformation potential (ADP), ionized impurities (IMP), and polar optical phonons (POP). The resulting carrier relaxation time can be calculated by Matthiessen’s rule\cite{dugdale1967mathiessen}:
    \begin{equation}
	\mathrm{\frac{1}{\tau}} = 	\mathrm{\frac{1}{\tau^{ADP}}} + 	\mathrm{\frac{1}{\tau^{IMP}}}  +	\mathrm{\frac{1}{\tau^{POP}}} 
    \end{equation}
    where $\mathrm{\tau^{ADP}}$, $\mathrm{\tau^{IMP}}$ and $\mathrm{\tau^{POP}}$ are the relaxation times resulting from ADP, IMP and POP scattering, respectively. A 77 $\times$ 77 $\times$ 77 dense $k$-grid is adopted to calculate carrier relaxation time and electron transport properties. Meanwhile, since the PBEsol usually underestimates the band gap and overestimates high-frequency dielectric constant, which is critical for POP calculations~\cite{xiong2025Chalcopyrite}, we have employed the Heyd-Scuseria-Ernzerhof (HSE06)~\cite{10.1063/1.1564060} hybrid functional for band gap and high-frequency dielectric constants calculations.


\clearpage

\section{Acknowledgements}
We would like to acknowledge Xiankun Zhang and Zhen Zhang for insightful discussions. We acknowledge the support received from the National Natural Science Foundation of China (Grant No. 12374024) and interdisciplinary research project for young teachers of USTB (Fundamental Research Funds for the Central Universities) (Grant No. FRF-IDRY-23-038). We also appreciate the computing resources provided by the USTB MatCom at the Beijing Advanced Innovation Center for Materials Genome Engineering.

\section*{Conflict of Interest}

The authors declare no conflict of interest.


\setlength{\bibsep}{0.0cm}
\bibliographystyle{Wiley-chemistry}
	\bibliography{X2CuZ}

\begin{thebibliography}{10}

\bibitem{liang2025stable}
Y.~Liang, Y.~Huang, Z.~Chen, C.~Gao, L.~Wang, \emph{Journal of Materials
  Chemistry A} \textbf{2025}.

\bibitem{hu2025all}
J.~Hu, Y.~Sun, H.~Wu, Z.~Yu, J.~Zhu, F.~Guo, Z.~Liu, W.~Cai, J.~Sui,
  \emph{Advanced Functional Materials} \textbf{2025}, \emph{35}, 2418244.

\bibitem{chen2022thermoelectric}
Z.-G. Chen, W.-D. Liu, \emph{Journal of Materials Science \& Technology}
  \textbf{2022}, \emph{121}, 256.

\bibitem{solidstatephysics}
N.~W. Ashcroft, N.~D. Mermin, \emph{Solid State Physics}, Philadelphia:
  Suanders College \textbf{1979}.

\bibitem{nolas1999skutterudites}
G.~Nolas, D.~Morelli, T.~M. Tritt, \emph{Annual Review of Materials Science}
  \textbf{1999}, \emph{29}, 89.

\bibitem{zhang2018deep}
Q.~Zhang, Q.~Song, X.~Wang, J.~Sun, Q.~Zhu, K.~Dahal, X.~Lin, F.~Cao, J.~Zhou,
  S.~Chen, et~al., \emph{Energy \& Environmental Science} \textbf{2018},
  \emph{11}, 933.

\bibitem{pei2014high}
Y.-L. Pei, H.~Wu, D.~Wu, F.~Zheng, J.~He, \emph{Journal of the American
  Chemical Society} \textbf{2014}, \emph{136}, 13902.

\bibitem{liu2025realizing}
K.~Liu, C.~Chen, J.~Cheng, X.~Ma, J.~Li, X.~Bao, H.~Li, Q.~Zhang, Y.~Chen,
  \emph{Advanced Functional Materials} \textbf{2025}, \emph{35}, 2419145.

\bibitem{zhang2021band}
X.~Zhang, Z.~Wang, B.~Zou, M.~K. Brod, J.~Zhu, T.~Jia, G.~Tang, G.~J. Snyder,
  Y.~Zhang, \emph{Chemistry of Materials} \textbf{2021}, \emph{33}, 9624.

\bibitem{pei2012band}
Y.~Pei, H.~Wang, G.~J. Snyder, \emph{Advanced Materials} \textbf{2012},
  \emph{24}, 6125.

\bibitem{guo2023enhanced}
Z.~Guo, G.~Wu, X.~Tan, R.~Wang, Z.~Zhang, G.~Wu, Q.~Zhang, J.~Wu, G.-Q. Liu,
  J.~Jiang, \emph{Advanced Functional Materials} \textbf{2023}, \emph{33},
  2212421.

\bibitem{liu2012convergence}
W.~Liu, X.~Tan, K.~Yin, H.~Liu, X.~Tang, J.~Shi, Q.~Zhang, C.~Uher,
  \emph{Physical Review Letters} \textbf{2012}, \emph{108}, 166601.

\bibitem{zheng2022synergistically}
J.~Zheng, T.~Hong, D.~Wang, B.~Qin, X.~Gao, L.-D. Zhao, \emph{Acta Materialia}
  \textbf{2022}, \emph{232}, 117930.

\bibitem{B916400F}
M.~Christensen, S.~Johnsen, B.~B. Iversen, \emph{Dalton Trans.} \textbf{2010},
  \emph{39}, 978.

\bibitem{https://doi.org/10.1002/idm2.12134}
H.~Xie, L.-D. Zhao, M.~G. Kanatzidis, \emph{Interdisciplinary Materials}
  \textbf{2024}, \emph{3}, 5.

\bibitem{ren2016contribution}
G.-K. Ren, J.-L. Lan, K.~J. Ventura, X.~Tan, Y.-H. Lin, C.-W. Nan, \emph{Npj
  Computational Materials} \textbf{2016}, \emph{2}, 1.

\bibitem{zhao2017defect}
M.~Zhao, W.~Pan, C.~Wan, Z.~Qu, Z.~Li, J.~Yang, \emph{Journal of the European
  Ceramic Society} \textbf{2017}, \emph{37}, 1.

\bibitem{https://doi.org/10.1002/adfm.202108532}
J.~He, Y.~Xia, W.~Lin, K.~Pal, Y.~Zhu, M.~G. Kanatzidis, C.~Wolverton,
  \emph{Advanced Functional Materials} \textbf{2022}, \emph{32}, 2108532.

\bibitem{https://doi.org/10.1002/advs.202417292}
Z.~Xia, X.~Shen, J.~Zhou, Y.~Huang, Y.~Yang, J.~He, Y.~Xia, \emph{Advanced
  Science} \textbf{2025}, \emph{12}, 2417292.

\bibitem{fu2014high}
C.~Fu, T.~Zhu, Y.~Pei, H.~Xie, H.~Wang, G.~J. Snyder, Y.~Liu, Y.~Liu, X.~Zhao,
  \emph{Advanced Energy Materials} \textbf{2014}, \emph{4}, 1400600.

\bibitem{2011Convergence}
Y.~Pei, X.~Shi, LaLonde, Aaron, H.~Wang, L.~Chen, Snyder, G.~Jeffrey,
  \emph{Nature} \textbf{2011}.

\bibitem{ti2022thermoelectric}
Z.~Ti, S.~Guo, X.~Zhang, J.~Li, Y.~Zhang, \emph{Journal of Materials Chemistry
  A} \textbf{2022}, \emph{10}, 5593.

\bibitem{he2019designing}
J.~He, Y.~Xia, S.~S. Naghavi, V.~Ozoli{\c{n}}{\v{s}}, C.~Wolverton,
  \emph{Nature Communications} \textbf{2019}, \emph{10}, 719.

\bibitem{xiong2025forbidden}
W.~Xiong, Z.~Han, Z.~Xia, Z.~Yang, J.~He, \emph{arXiv preprint
  arXiv:2507.01256} \textbf{2025}.

\bibitem{https://doi.org/10.1002/anie.201508381}
W.~G. Zeier, A.~Zevalkink, Z.~M. Gibbs, G.~Hautier, M.~G. Kanatzidis, G.~J.
  Snyder, \emph{Angewandte Chemie International Edition} \textbf{2016},
  \emph{55}, 6826.

\bibitem{PhysRevB.55.13605}
S.-H. Wei, A.~Zunger, \emph{Phys. Rev. B} \textbf{1997}, \emph{55}, 13605.

\bibitem{C4EE00997E}
L.-D. Zhao, J.~He, D.~Berardan, Y.~Lin, J.-F. Li, C.-W. Nan, N.~Dragoe,
  \emph{Energy Environ. Sci.} \textbf{2014}, \emph{7}, 2900.

\bibitem{doi:10.1021/jacs.1c10284}
C.~Zhang, J.~He, R.~McClain, H.~Xie, S.~Cai, L.~N. Walters, J.~Shen, F.~Ding,
  X.~Zhou, C.~D. Malliakas, J.~M. Rondinelli, M.~G. Kanatzidis, C.~Wolverton,
  V.~P. Dravid, K.~R. Poeppelmeier, \emph{Journal of the American Chemical
  Society} \textbf{2022}, \emph{144}, 2569, pMID: 35113569.

\bibitem{https://doi.org/10.1002/adma.202104908}
W.~Lin, J.~He, X.~Su, X.~Zhang, Y.~Xia, T.~P. Bailey, C.~C. Stoumpos, G.~Tan,
  A.~J.~E. Rettie, D.~Y. Chung, V.~P. Dravid, C.~Uher, C.~Wolverton, M.~G.
  Kanatzidis, \emph{Advanced Materials} \textbf{2021}, \emph{33}, 2104908.

\bibitem{doi:10.1021/jacs.4c16394}
A.~Bhui, P.~V.~D. Matukumilli, S.~Biswas, A.~Ahad, A.~Ghata, D.~Rawat,
  M.~Dutta, A.~Soni, U.~V. Waghmare, K.~Biswas, \emph{Journal of the American
  Chemical Society} \textbf{2025}, \emph{147}, 3758, pMID: 39807771.

\bibitem{PhysRev.80.72}
J.~Bardeen, W.~Shockley, \emph{Physical Review} \textbf{1950}, \emph{80}, 72.

\bibitem{PhysRevX.15.011066}
Y.~Wang, L.~Xie, H.~Yang, M.~Hu, X.~Qian, R.~Yang, J.~He, \emph{Physical Review
  X} \textbf{2025}, \emph{15}, 011066.

\bibitem{kim2021extremely}
S.~E. Kim, F.~Mujid, A.~Rai, F.~Eriksson, J.~Suh, P.~Poddar, A.~Ray, C.~Park,
  E.~Fransson, Y.~Zhong, et~al., \emph{Nature} \textbf{2021}, \emph{597}, 660.

\bibitem{10.1063/1.4904513}
J.~Liu, G.-M. Choi, D.~G. Cahill, \emph{Journal of Applied Physics}
  \textbf{2014}, \emph{116}, 233107.

\bibitem{mues1980na2auas}
C.~Mues, H.-U. Schuster, \emph{Zeitschrift f{\"u}r Naturforschung B}
  \textbf{1980}, \emph{35}, 1055.

\bibitem{https://doi.org/10.1002/zaac.200900417}
S.-J. Kim, G.~Miller, J.~Corbett, \emph{Zeitschrift f{\"u}r anorganische und
  allgemeine Chemie} \textbf{2010}, \emph{636}, 67.

\bibitem{kirklin2015open}
S.~Kirklin, J.~E. Saal, B.~Meredig, A.~Thompson, J.~W. Doak, M.~Aykol,
  S.~R{\"u}hl, C.~Wolverton, \emph{Npj Computational Materials} \textbf{2015},
  \emph{1}, 1.

\bibitem{bronger1992synthese}
W.~Bronger, H.~Kathage, \emph{Journal of Alloys and Compounds} \textbf{1992},
  \emph{184}, 87.

\bibitem{strahle1974kristalldaten}
J.~Str{\"a}hle, K.-P. L{\"o}rcher, \emph{Zeitschrift f{\"u}r Naturforschung B}
  \textbf{1974}, \emph{29}, 266.

\bibitem{jagodzinski1959kristallstruktur}
H.~Jagodzinski, \emph{Zeitschrift f{\"u}r Kristallographie-Crystalline
  Materials} \textbf{1959}, \emph{112}, 80.

\bibitem{havinga1972compounds}
E.~Havinga, H.~Damsma, P.~Hokkeling, \emph{Journal of the Less Common Metals}
  \textbf{1972}, \emph{27}, 169.

\bibitem{pyykko2009molecular}
P.~Pyykk{\"o}, M.~Atsumi, \emph{Chemistry--A European Journal} \textbf{2009},
  \emph{15}, 186.

\bibitem{contreras2011nciplot}
J.~Contreras-Garc{\'\i}a, E.~R. Johnson, S.~Keinan, R.~Chaudret, J.-P.
  Piquemal, D.~N. Beratan, W.~Yang, \emph{Journal of chemical theory and
  computation} \textbf{2011}, \emph{7}, 625.

\bibitem{doi:10.1021/j150495a016}
L.~Pauling, \emph{The Journal of Physical Chemistry} \textbf{1952}, \emph{56},
  361.

\bibitem{cui1990bonding}
C.~Cui, M.~Kertesz, \emph{Inorganic Chemistry} \textbf{1990}, \emph{29}, 2568.

\bibitem{li2021demonstration}
A.~Li, C.~Hu, B.~He, M.~Yao, C.~Fu, Y.~Wang, X.~Zhao, C.~Felser, T.~Zhu,
  \emph{Nature Communications} \textbf{2021}, \emph{12}, 5408.

\bibitem{akhmedova2009effect}
G.~Akhmedova, D.~S. Abdinov, \emph{Inorganic Materials} \textbf{2009},
  \emph{45}, 854.

\bibitem{zhang2020promising}
W.~Zhang, C.~Chen, H.~Yao, W.~Xue, S.~Li, F.~Bai, Y.~Huang, X.~Li, X.~Lin,
  F.~Cao, et~al., \emph{Chemistry of Materials} \textbf{2020}, \emph{32}, 6983.

\bibitem{tritt2005thermal}
T.~M. Tritt, \emph{Thermal conductivity: theory, properties, and applications},
  Springer Science \& Business Media \textbf{2005}.

\bibitem{10.1063/1.1564060}
J.~Heyd, G.~E. Scuseria, M.~Ernzerhof, \emph{The Journal of Chemical Physics}
  \textbf{2003}, \emph{118}, 8207.

\bibitem{ganose2021efficient}
A.~M. Ganose, J.~Park, A.~Faghaninia, R.~Woods-Robinson, K.~A. Persson,
  A.~Jain, \emph{Nature Communications} \textbf{2021}, \emph{12}, 2222.

\bibitem{han2025weak}
Z.~Han, J.~Wang, C.~Zhang, Z.~Xia, J.~He, \emph{Inorganic Chemistry}
  \textbf{2025}.

\bibitem{doi:10.1073/pnas.1711725114}
J.~Mao, J.~Shuai, S.~Song, Y.~Wu, R.~Dally, J.~Zhou, Z.~Liu, J.~Sun, Q.~Zhang,
  C.~dela Cruz, S.~Wilson, Y.~Pei, D.~J. Singh, G.~Chen, C.-W. Chu, Z.~Ren,
  \emph{Proceedings of the National Academy of Sciences} \textbf{2017},
  \emph{114}, 10548.

\bibitem{vasp1}
G.~Kresse, J.~Furthm{\"u}ller, \emph{Physical Review B} \textbf{1996},
  \emph{54}, 11169.

\bibitem{vasp2}
G.~Kresse, J.~Furthm{\"u}ller, \emph{Computational Materials Science}
  \textbf{1996}, \emph{6}, 15.

\bibitem{paw1}
P.~E. Bl{\"o}chl, \emph{Physical Review B} \textbf{1994}, \emph{50}, 17953.

\bibitem{paw2}
G.~Kresse, D.~Joubert, \emph{Physical Review B} \textbf{1999}, \emph{59}, 1758.

\bibitem{pbesol1}
G.~I. Csonka, J.~P. Perdew, A.~Ruzsinszky, P.~H. Philipsen, S.~Leb{\`e}gue,
  J.~Paier, O.~A. Vydrov, J.~G. {\'A}ngy{\'a}n, \emph{Physical Review B}
  \textbf{2009}, \emph{79}, 155107.

\bibitem{nelson2020lobster}
R.~Nelson, C.~Ertural, J.~George, V.~L. Deringer, G.~Hautier, R.~Dronskowski,
  \emph{Journal of Computational Chemistry} \textbf{2020}, \emph{41}, 1931.

\bibitem{mostofi2008wannier90}
A.~A. Mostofi, J.~R. Yates, Y.-S. Lee, I.~Souza, D.~Vanderbilt, N.~Marzari,
  \emph{Computer physics communications} \textbf{2008}, \emph{178}, 685.

\bibitem{zhang2022magnetictb}
Z.~Zhang, Z.-M. Yu, G.-B. Liu, Y.~Yao, \emph{Computer Physics Communications}
  \textbf{2022}, \emph{270}, 108153.

\bibitem{giannozzi2017advanced}
P.~Giannozzi, O.~Andreussi, T.~Brumme, O.~Bunau, M.~B. Nardelli, M.~Calandra,
  R.~Car, C.~Cavazzoni, D.~Ceresoli, M.~Cococcioni, et~al., \emph{Journal of
  physics: Condensed matter} \textbf{2017}, \emph{29}, 465901.

\bibitem{phonopy-phono3py}
A.~Togo, \emph{Journal of the Physical Society of Japan} \textbf{2023},
  \emph{92}, 012001.

\bibitem{zhou2014lattice}
F.~Zhou, W.~Nielson, Y.~Xia, V.~Ozoli{\c{n}}{\v{s}}, \emph{Physical Review
  Letters} \textbf{2014}, \emph{113}, 185501.

\bibitem{werthamer1970self}
N.~Werthamer, \emph{Physical Review B} \textbf{1970}, \emph{1}, 572.

\bibitem{han2022fourphonon}
Z.~Han, X.~Yang, W.~Li, T.~Feng, X.~Ruan, \emph{Computer Physics
  Communications} \textbf{2022}, \emph{270}, 108179.

\bibitem{dugdale1967mathiessen}
J.~Dugdale, Z.~Basinski, \emph{Physical Review} \textbf{1967}, \emph{157}, 552.

\bibitem{xiong2025Chalcopyrite}
Z.~H. D. Y. J.~H. W.~Xiong, Z.~Xia, \emph{arXiv preprint arXiv:2508.08988}
  \textbf{2025}.

\end{thebibliography}






\end{document}